\documentclass[fleqn,10pt]{wlscirep}
\usepackage[utf8]{inputenc}
\usepackage[T1]{fontenc}
\usepackage{svg}
\usepackage[clean]{svg-extract}
\usepackage{nameref}

\newcommand*\samethanks[1][\value{footnote}]{\footnotemark[#1]}

\title{Enhancing Drug Discovery: Quantum Machine Learning for QSAR Prediction with Incomplete Data}

\author[1]{Wei-Yin Chiang\thanks{This work was conducted while the author was employed at Insilico Medicine Taiwan Ltd.}}
\author[1]{Po-Yu Kao\samethanks[1]}
\author[1]{Tzu-Lan Yeh}
\author[1]{Ya-Chu Yang\samethanks[1]}
\author[1,2]{Yen-Chu Lin\samethanks[1]}
\author[3,$\dagger$]{Alex Zhavoronkov}

\affil[1]{Insilico Medicine Taiwan Ltd., Taipei, 110208, Taiwan}
\affil[2]{Department of Pharmacy, National Yang Ming Chiao Tung University, Taipei 112304, Taiwan}
\affil[3]{Insilico Medicine Hong Kong Ltd., Hong Kong SAR, 999077, China}
\affil[$\dagger$]{alex@insilicomedicine.com}

\begin{abstract}

Qualitative structure-activity relationship (QSAR) is important for drug discovery and offers valuable insights into the biological interactions of potential drug candidates. It has been demonstrated that QSAR can be accurately predicted by machine learning. However, data with poor quality and limited availability are always the most common and critical issues for medical-related applications for machine learning.
In this manuscript, we aim to discuss the performance of classical and quantum classifiers in QSAR prediction and attempt to demonstrate the quantum advantages in the generalization power of the quantum classifier under conditions of limited data availability and a reduced number of features. By applying different data embedding methods followed by feature selection through principal component analysis (PCA), we find that the quantum classifier outperforms the classical one when a small number of features are selected and the number of training samples is limited. The generality of quantum advantages in other open datasets is also explored.

\end{abstract}
\begin{document}

\flushbottom
\maketitle
%vvv
\thispagestyle{empty}

\section*{Introduction}
% importance of QSAR --> Ken
In the realm of modern drug discovery, the perpetual challenge lies in developing new pharmaceutical compounds that demonstrate optimal therapeutic efficacy while minimizing side effects. Quantitative structure-activity relationship (QSAR) analysis has emerged as a pivotal computational tool to tackle this obstacle, bridging the gap between compounds' molecular structure and biological activity. By establishing quantitative correlations between chemical structures and their corresponding bioactivities, QSAR models offer valuable insights into the biological interactions of potential drug candidates. This, in turn, expedites the screening of vast chemical libraries, facilitating the drug discovery process and reducing associated costs. Nevertheless, with the ever-increasing complexity of molecular systems and the demand for higher prediction accuracy, novel approaches are continuously sought to meet the field's needs. 

Over the years, QSAR has revolutionized drug discovery, guiding the design and development of effective drugs and significantly reducing the time and cost associated with traditional trial-and-error approaches.
% classical QSAR --> Ken
Hansch and Fujita \cite{hansch1964p} introduced the concept of QSAR and laid the foundation for the field. 
They proposed a quantitative method, known as $\rho$-$\sigma$-$\pi$ analysis, for correlating biological activity with chemical structure based on electronic and steric parameters. This pioneering study demonstrated the potential of QSAR in guiding the rational design of bioactive molecules, revolutionizing the drug discovery process. Zhou et al. \cite{zhou2006boosting} proposed and validated the boosting support vector regression (BSVR) method, which combines boosting and SVR to yield improved results in QSAR studies compared to traditional regression techniques like multiple linear regression (MLR) and conventional support vector regression (SVR). Pourbasheer et al. \cite{pourbasheer2009application} introduced the support vector machine (SVM) as an effective tool for QSAR modeling, the utilization of genetic algorithm (GA) for feature selection, and providing valuable insights into the key descriptors influencing the Ca$^{2+}$-activated K$^{+}$ (BK)-channel activity of the compounds studied. These findings contribute to the field of computational drug discovery and can have practical implications in designing new BK-channel activators with desired biological activities.

The work of Polishchuk et al. \cite{polishchuk2009application} includes the application of random forest (RF) to QSAR analysis for aquatic toxicity, the use of simplex descriptors, successful model development and validation, identification of key descriptors affecting toxicity, demonstration of model robustness, and the proposal of a fast optimization procedure. Their work advances the understanding and practical applications of QSAR modeling for assessing the aquatic toxicity of chemical compounds. Dahl et al. \cite{dahl2014multi} introduced a novel approach for multi-assay prediction using artificial neural networks (ANN), leveraging recent methods to address overfitting, and demonstrating superior performance compared to alternative methods. These findings contribute to the evolving landscape of quantitative structure-activity/property relationship (QSAR/QSPR) research and open up new possibilities for using neural networks to tackle complex predictive modeling tasks in chemistry and drug design. Kwon et al. \cite{kwon2019comprehensive} introduced a comprehensive ensemble method with second-level meta-learning, the development of an end-to-end neural network-based individual classifier for sequential feature extraction from SMILES representations, and the successful application of these methods to improve predictive performance in the QSAR prediction task. Their findings advance the field of ensemble learning and demonstrate the practical applicability of neural network-based classifiers in QSAR studies.
Pourbasheer et al. \cite{pourbasheer2019qsar} successfully developed and evaluated QSAR models for predicting casein kinase 2 (CK2) inhibitor activity using MLR and SVM methods. The use of feature selection with the GA, identification of important molecular descriptors, and the reported high statistical parameters add to the significance of this research. The results can have practical implications in the design of new and potent CK2 inhibitors for potential therapeutic applications.
Hu et al. \cite{hu2020deep} introduced a novel deep learning-based method for QSAR prediction, the effective representation of chemical molecules through the encoder-decoder model, the robust convolutional neural network (CNN) framework for prediction, and the demonstration of improved performance compared to state-of-the-art methods. The findings of this research have practical relevance in drug discovery and provide valuable insights into leveraging deep learning techniques for QSAR analysis.

Kim et al. \cite{kim2020extension} proposed and evaluated an extended model of Profile-QSAR (pQSAR) 2.0 that combines random forest regressors (RFRs) and partial least squares regressors (PLSRs) to improve the predictive performance of QSAR regression models. The iterative updating process and the ensemble approach enhance the accuracy and reliability of the predictions. The findings of this research have potential applications in drug discovery and computational chemistry, where QSAR models are widely used to guide compound screening and optimization efforts.
The collective contributions of these studies demonstrate the continual advancements and diversification of machine learning and deep learning techniques in QSAR modeling.  

%quantum QSAR
QSAR can also be predicted using parameterized quantum circuit (PQC)-based quantum machine learning. PQC consists of quantum bits, rotation gates, and measurements \cite{11benedetti2019parameterized}. The learnable parameters in this quantum circuit control the rotation angles of the quantum gates and are updated by minimizing the cost function estimated classically. 
PQC can collaborate with the classical network to form a hybrid quantum-classical network. 
Several applications have been demonstrated \cite{11benedetti2019parameterized,15biamonte2017quantum}, including figure generation \cite{16tsang2022hybrid,17huang2021experimental}, data regression \cite{4suzuki2020predicting}, and classification \cite{10wu2021application,14havlivcek2019supervised}. Due to the larger Hilbert space inherited from the fundamental properties of quantum mechanics, quantum circuits often outperform classical models, a phenomenon known as quantum advantage, which has been demonstrated in various aspects. Most research shows that the hybrid model has lower model complexity, higher training efficiency \cite{7batra2021quantum,8liu2021rigorous}, greater expressive power \cite{18du2020expressive}, robustness to noise, and better overall performance \cite{8liu2021rigorous}. Quantum advantages of the hybrid model have also been found in drug discovery \cite{12cao2018potential}, where ligands generated from quantum generative adversarial networks (GAN) \cite{5kao2023exploring} demonstrate better drug-likeness compared to those generated using classical GANs. 

In QSAR prediction, the hybrid quantum-classical convolutional neural network exhibits lower complexity, a 20\% reduction when applied to protein-ligand binding affinity prediction, accompanied by a 40\% reduction in the training process time while maintaining the same level of performance \cite{3domingo2023hybrid}. 
For drug response $IC_{50}$ prediction, the hybrid quantum model's effectiveness value outperforms the classical one by 15\% in test loss \cite{2sagingalieva2023hybrid}. Additionally, the quantum-classical model demonstrates greater expressive power in toxicity prediction \cite{4suzuki2020predicting}.
However, data with poor quality and limited availability in medical applications are consistently the most frequent and severe issues for machine learning \cite{19hekler2019we}. This also becomes a legal concern when seeking approval for commercial usage in medical devices.

In this manuscript, we aim to discuss the performance of pure classical and hybrid quantum-classical networks, denoted by classical and quantum classifiers, respectively, on QSAR prediction and demonstrate the quantum advantages in the generalization power of the hybrid model, especially in scenarios with limited data availability and feature numbers. When we apply various embedding methods to the data and perform feature selection through principal component analysis (PCA) \cite{jolliffe2002principal}, we observe that the quantum classifier outperforms its classical counterpart, especially when only a few features are considered. The generality of the quantum advantage in other open datasets is further discussed.

\section*{Methods}

\begin{figure}[h!]
    \centering
    \includegraphics[width=0.9\linewidth]{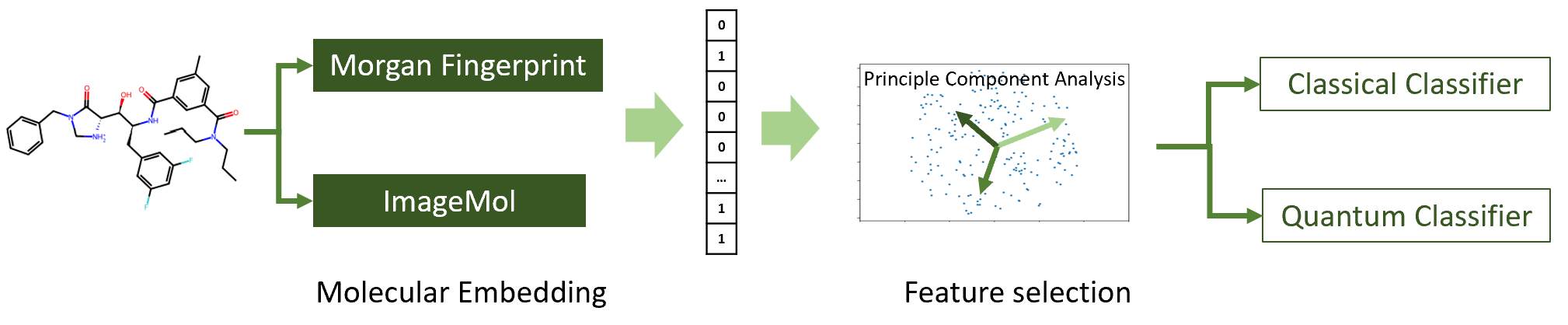}
    \caption{The proposed workflow. The inhibitor structure(SMILES) is embedded by Morgan fingerprint or ImageMol followed by feature selection (PrincipleComponent Analysis, PCA). $2^n$ selected features are fed into classical and quantum classifiers. Where $n$ is the number of qubits in the quantum classifier.}
    \label{fig:workflow}
\end{figure}

Fig.\ref{fig:workflow} illustrates the workflow. In our proposed quantum QSAR model, input molecules are first encoded by Morgan \cite{rogers2010extended} fingerprints and ImageMol \cite{zeng2022accurate} embeddings. PCA is then applied for dimensionality reduction and tries to mimic the data incompleteness. In these experiments, $2^n$ features are used for classification, where $n$ is the number of the qubits. Finally, the reduced embedding is fed into classical or quantum classifiers for performance comparison of the classifiers. In the following content, we will introduce the methods of embedding, feature selection and the architecture of classical/quantum classifiers used in this work.

\subsection*{Molecular Embedding}\label{subsection:molecular_embedding} % Ken
In this section, we will provide a concise overview of the two distinct types of molecular embeddings utilized in our study: local-structure-based \cite{rogers2010extended}, and image-based \cite{zeng2022accurate} molecular embeddings. Please refer to their respective papers for further in-depth information on these embeddings.

\subsubsection*{Morgan Fingerprint} % Ken
Morgan fingerprint, also known as Extended-Connectivity Circular Fingerprints (ECFP) \cite{rogers2010extended}, is a type of molecular fingerprint commonly used in cheminformatics and drug discovery. 
The Morgan algorithm encodes molecular structures into binary bit strings, representing a molecule's presence or absence of certain substructures or molecular features.
The Morgan fingerprint is generated by iteratively expanding from each atom in the molecule, considering its neighboring atoms up to a specified radius. 
The neighbors' identities are incorporated into the fingerprint at each iteration, including more distant information from the original atom. 
This iterative process continues until the specified radius is reached, creating a circular fingerprint representing each atom's local chemical environment. 
In this work, RDKit \cite{greg_landrum_2022_7179566} is used with default radius and 512 bit to extract the Morgan fingerprint for each molecule. 

%\subsubsection*{molecularGNN} % Ken
%Tsubaki et al. \cite{tsubaki2019compound} employ end-to-end representation learning through deep neural networks to predict compound-protein interactions (CPIs) using discrete symbolic data, such as graphs for compounds and protein sequences. Compounds are represented as graphs, with atoms as vertices and chemical bonds as edges. Proteins, on the other hand, are represented as sequences with amino acids as characters. They utilize a graph neural network (GNN) to capture compound representations, while a convolutional neural network (CNN) is employed for protein representations. These representations are integrated to establish an end-to-end learning system for CPIs prediction. In our study, using benchmarking datasets, we first pretrain their classical GNN \footnote{\url{https://github.com/masashitsubaki/molecularGNN_smiles/tree/master}}. The trained GNNs are then used as feature extractors to obtain graph embeddings for each molecule. These extracted molecular vectors are referred to as \textit{molecularGNN}.

\subsubsection*{ImageMol} % Ken
ImageMol \cite{zeng2022accurate} presents a powerful deep-learning framework for computational drug discovery. 
Its ability to learn from molecular images in an unsupervised manner and its high performance in predicting various molecular properties and target profiles make it a promising tool for identifying potential drug candidates, including anti-SARS-CoV-2 molecules, for possible treatment of COVID-19.
The framework of ImageMol combines image processing techniques with comprehensive molecular chemistry knowledge to extract fine pixel-level molecular features in a visual computing approach.
It offers two significant improvements over existing methods:
\begin{itemize}
  \item \textbf{Molecular Image Representation:} ImageMol uses molecular images as the feature representation of compounds. This means that instead of traditional molecular representations like SMILES strings or molecular fingerprints, the compounds are represented as images, which helps achieve higher accuracy in predictions with relatively lower computing costs.
  \item \textbf{Unsupervised Pretrained Learning:} ImageMol employs an unsupervised pretrained learning approach, where it learns to capture structural information from a vast dataset of 10 million drug-like compounds. The dataset covers diverse biological activities at the human proteome, allowing the framework to generalize well across different drug discovery tasks.
\end{itemize}

Pretrained ImageMol \footnote{\url{https://hub.docker.com/r/pykao/imagemol}} is used to extract 512-dimensional embedding for each molecule.  

\subsection*{Feature Selection} \label{subsection: PCA}
Principal component analysis (PCA) \cite{jolliffe2002principal} is a standard statistical method used in machine learning for the reduction of feature dimension, visualization, and feature extraction. It can be applied to a dataset of $L$ data points with $M$ dimensional features, represented as D $\in \mathbb{R}^{L \times M}$. The goal is to determine the principal components of the data by calculating the eigenvalues and eigenvectors of the covariance matrix, which is a $M\times M$ matrix showing the covariances between features and project the features on these principal components.  After performing PCA, we can reduce the dimensionality of the original features by selecting principal components with corresponding top k largest eigenvalues ($k<M$). In our work, we extract the top-$N$ components for the classifier to mimic the incomplete data feature.

\subsection*{Classifier}
\begin{figure}[h!]
    \centering
    \includegraphics[width=0.8\linewidth]{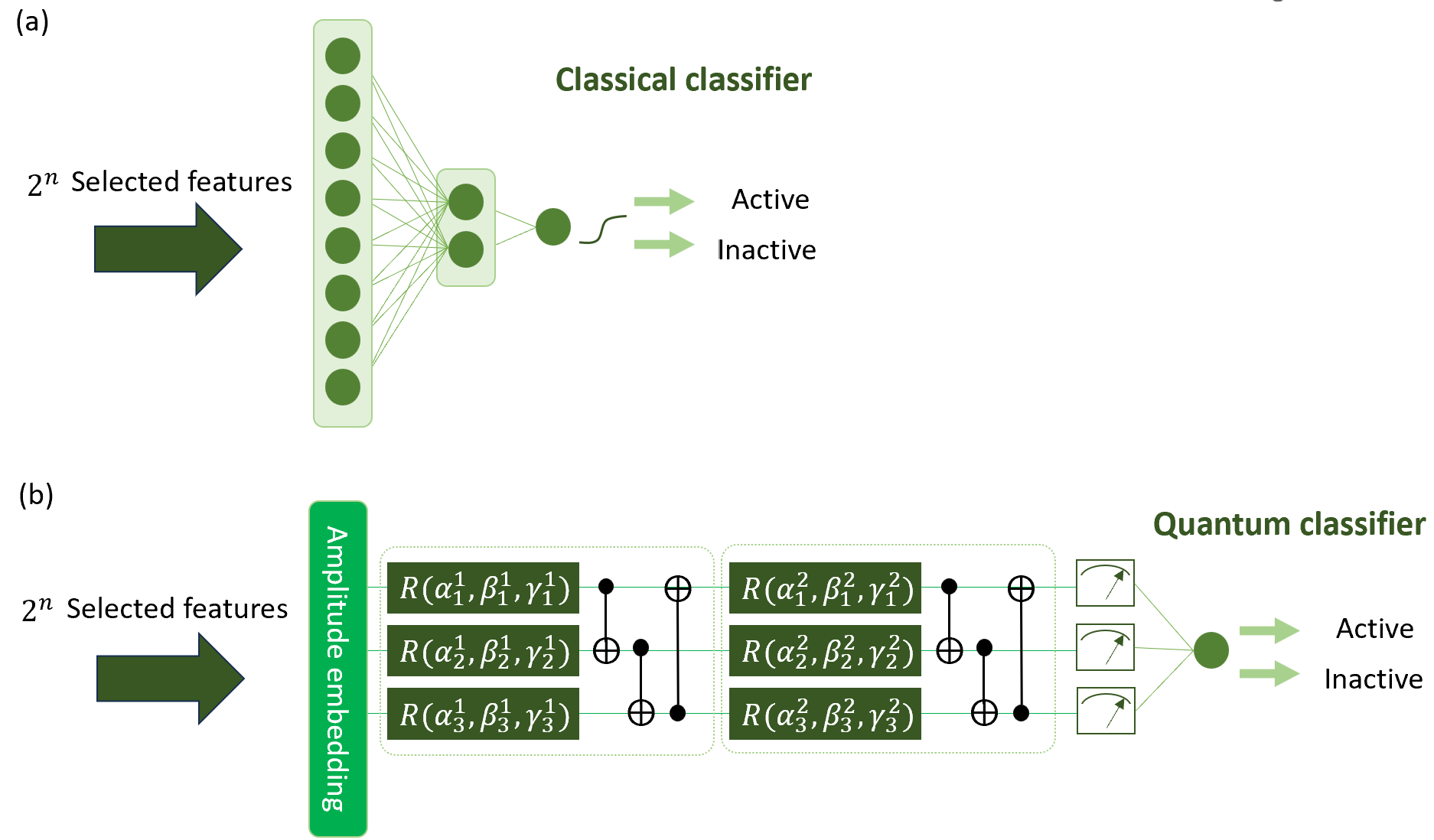}%{Figures/quantum_qsar_workflow.svg}
    \caption{The architectural of classifiers. (a) In classical classifier, the classical multi-layer perceptron (MLP) has an input layer with $2^n$ neurons, a hidden layer with two nodes, and an output node. (b) The quantum classifier consists of an amplitude encoding layer to encode input with the dimension of $2^n$, two strongly entangling layers followed by the measurements to all qubits. The measured results are then aggregated into an output.}
    \label{fig:classifier}
\end{figure}

To assure the simplicity and generality of the classical classifier, the architecture of a three-layer perceptron with the dimension of $N \times 2 \times 1$ is considered for the classical classifier and is depicted in Fig.\ref{fig:classifier}(a) and the number of the trainable parameter is $2(N+1)$. Where $N$ is the feature number.

The quantum classifier consists of the parameterized quantum circuit (PQC) made by $n$ qubits followed by a trainable ansatz and a linear regression function (Fig.\ref{fig:classifier}(b), where $n$ is $\log_2(N)$. Two strongly entangled layers construct the ansatz of the circuit, and each layer has $3 \times n$ variables. When the selected features are first encoded to the quantum states via the amplitude embedding method, the ansatz operates the quantum states followed by the measurement to all qubits. After measuring all qubits, the regression function makes the binary prediction with a zero threshold. The number of trainable parameters is $7n$. 

$80\%$ and $20\%$ of the whole data are set for training and testing data(train-test split), respectively. The accuracy of the classifiers is determined by (i) finding the best performance of each training over $100$ epoch, (ii)calculating the mean of best accuracy over the $20$ times' training, (iii)re-sample the training and testing set followed by (i) and (ii), repeat $5$ times, and take the average.

\section*{Results}

Many works focus on the quantum advantages of the model with complicated/sophisticated architecture, and it is challenging to compare classical and quantum models fairly under this scenario because the benefits might come from intractable resources originating from the complicated architecture both in quantum and classical classifiers. Therefore, characterizing the fundamental behavior of both classifiers with the most concise structure will give the intrinsic difference between quantum and classical scenes. In this section, we investigate the effect of feature numbers and embedding methods in classical and quantum classifiers and attempt to identify the quantum advantage in different scenarios. We identify the robustness of the quantum classifier as a quantum advantage. With this advantage, we find the performance of the quantum classifier degenerates with the reduction of training data, less than the classical classifier, in a low dimensional feature regime. Finally, we explore the generality of the quantum advantage on different datasets.

\subsection*{Quantum classifiers outperform classical classifiers when the number of selected features is small.}
Accuracy is used to estimate the model performances of classical and quantum classifiers. The data is embedded by Morgan fingerprint (MGFP) and ImageMol (IMGMOL), followed by feature selection obtained from PCA. The top-$2^n$ features are selected to test the performance of classifiers. We find the classical classifier performs better than the quantum one when the feature number is enormous. However, when the number of selected features is small, the quantum classifier outperforms the classical one. From the 
%Fig.\ref{fig:MGFP_IMGMOL} and 
Table.\ref{Tab:MGFP_IMGMOL}, we can find the classical classifier performs better when data embedded by MGFP with $2^8$ ($n=8$) features by approximately 5\%. On the other hand, the quantum classifier outperforms the classical one when $n=3$, a pronounced enhancement in accuracy is found by approximately 8\%.

%\begin{figure}[h!]
%    \centering
%    \includegraphics[width=0.6\linewidth]{Figures/embedding_dim.png}
%    \caption{Performances of the classical and quantum classifiers. The histogram %shows prediction accuracies under MGFP and IMGMOL embeddings for both $n=3$ and %$n=8$. The figure illustrates that the quantum classifier outperforms the classical %one when $n$ is small. (The BACE dataset is embedded using MGFP and IMGMOL.)}
%    \label{fig:MGFP_IMGMOL}
%\end{figure}

\begin{table}[]
\centering
\begin{tabular}{ccccc}
 & \multicolumn{2}{c}{MGFP} & \multicolumn{2}{c}{IMGMOL} \\
\hline
n & Classical    & Quantum   & Classical     & Quantum    \\
\hline
2 & $0.60\pm0.03$ &$0.66\pm0.03$ & $0.60\pm0.03$& $0.62\pm0.02$\\
3 & $0.69\pm0.02$ &$0.75\pm0.02$ & $0.65\pm0.04$& $0.70\pm0.02$\\
4 & $0.75\pm0.02$ &$0.75\pm0.02$ & $0.66\pm0.02$& $0.72\pm0.02$\\
8 & $0.80\pm 0.02$&$0.76\pm0.02$ &$0.80\pm0.02$ &$0.74\pm0.02$\\
\hline
\end{tabular}
\caption{The table shows the comparison of performances between classical and quantum classifiers on the data with different numbers of features and embedding methods. (BACE dataset)}
\label{Tab:MGFP_IMGMOL}
\end{table}

\subsection*{Quantum advantage found in low dimensional classifier appears in different embedding methods.}
In addition to the higher accuracy of the quantum classifier for low-dimensional data, Table \ref{Tab:MGFP_IMGMOL} also demonstrates that this superior performance is consistent across different data embeddings. Here, we compare the accuracies of the classical and quantum classifiers on data embedded by MGFP and IMGMOL, and we find that the performance of the quantum classifier is superior when a small number of features are selected in both embedding methods. Table \ref{Tab:MGFP_IMGMOL} also shows that when IMGMOL embeds the features, the accuracy of the quantum classifier slightly outperforms the classical one by approximately 8\% when $n=3$. Based on the results presented here, it is not evident which embedding method is suitable for any specific classifier. In the subsequent results, unless specified otherwise, we demonstrate the classification results of features embedded by MGFP on the BACE dataset.

\subsection*{The predictive ability of the quantum classifier shows robustness to the feature dimension.}
\begin{figure}[h!]
    \centering
    \includegraphics[width=\linewidth]{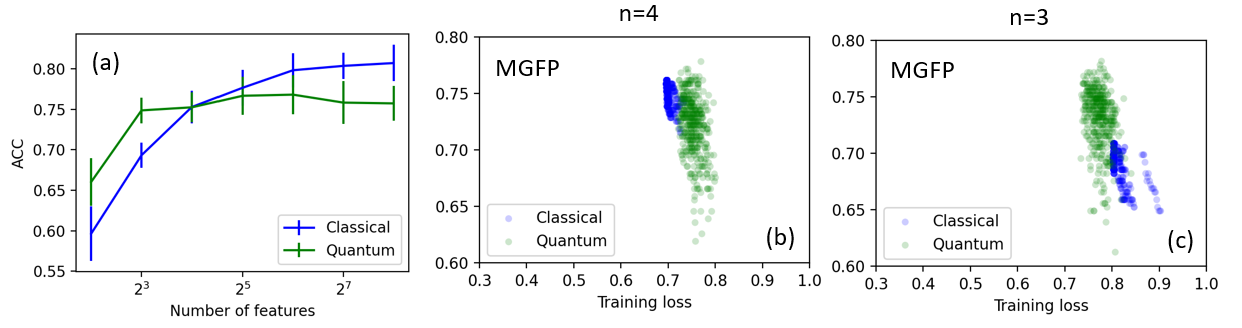}
    \caption{(a) The trend in accuracy with varying numbers of features. The change in accuracy is highly dependent on the number of features in the classical classifier. (b) and (c) show the relationships between training loss and testing accuracy for $n=4$ and $n=3$ respectively. From these training loss and testing accuracy results, it is evident that the quantum classifier exhibits superior generalization power compared to the classical classifier.}
    \label{fig:MGFP_ACC_n4_n8}
\end{figure}

To gain a deeper understanding of this quantum advantage, we systematically benchmark the performance of classical and quantum classifiers using MGFP-embedded features. The accuracies of classical and quantum classifiers for different feature numbers are depicted in Fig. \ref{fig:MGFP_ACC_n4_n8}(a). The figure illustrates that both classical and quantum classifiers exhibit a decrease in performance as the number of features decreases. This dependency is more pronounced in classical classifiers, leading to improved performance with high-dimensional features and a decline in performance with low dimensions, with a crossover point at $n=4$. In contrast, the performance of the quantum classifier is less affected by changes in feature dimension, suggesting that it is more robust to variations in feature number.

\subsection*{Quantum classifier shows comparable accuracy when the training loss is higher.}

To understand more about this quantum advantage, we focus on the crossover of the performances of classical and quantum classifiers, which occurs at $n=4$ in Fig. \ref{fig:MGFP_ACC_n4_n8}(a). We delve into the expressive and generalization power of each classifier by examining the relationship between training loss and testing accuracy. In Fig. \ref{fig:MGFP_ACC_n4_n8}(b), we observe that classical classifiers exhibit lower training loss but comparable testing accuracy to quantum classifiers. Additionally, upon analyzing the architecture of both classifiers, we find that the quantum classifier requires only 82\% of the trainable parameters of the classical classifier to achieve similar accuracies, highlighting the superior expressive power of the quantum classifier. Furthermore, when $n=3$ (as shown in Fig. \ref{fig:MGFP_ACC_n4_n8}(c)), in some cases, classical and quantum classifiers exhibit similar training loss but higher testing accuracy in quantum classifiers. This suggests that quantum classifiers possess better generalization power in scenarios involving low-dimensional features.

\subsection*{Quantum classifier performs better on the data with low-dimensional features for the small number of training samples.}

\begin{figure}[h!]
    \centering
    \includegraphics[width=0.9\linewidth]{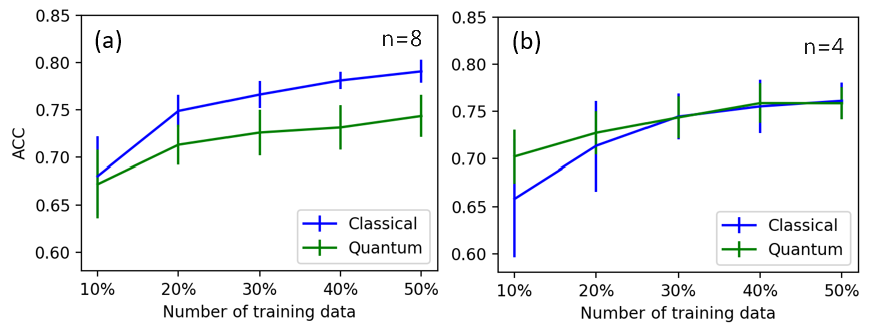}
    \caption{(a) and (b) are the dependencies of accuracies on the number of training data when $n=8$ and $n=4$, respectively. The figure shows that when $n=4$, the quantum advantage is found at the small number of training data regimes, while the classical classifier still outperforms the quantum one when $n=8$.}
    \label{fig:ACC_n4_n8_data_num}
\end{figure}

We benchmark the performance of classical and quantum classifiers using various numbers of training data. To examine the effect of the amount of training data on performance, we sampled 10\%, 20\%, 30\%, 40\%, and 50\% of the original whole training data into subgroups and trained the classifiers. In Fig. \ref{fig:ACC_n4_n8_data_num}, both classifiers' performances are improved as the number of training samples increases. While accuracy is significantly influenced by the number of training samples, the classical classifier consistently outperforms the quantum classifier when working with high-dimensional features (Fig. \ref{fig:ACC_n4_n8_data_num}(a)). However, this dependence leads to lower accuracy in the classical classifier when dealing with data represented by low-dimensional features in the small training data regime (Fig. \ref{fig:ACC_n4_n8_data_num}(b)).

\subsection*{Quantum classifiers show better generalization ability}
\begin{figure}[h!]
\includegraphics[width=1\linewidth]{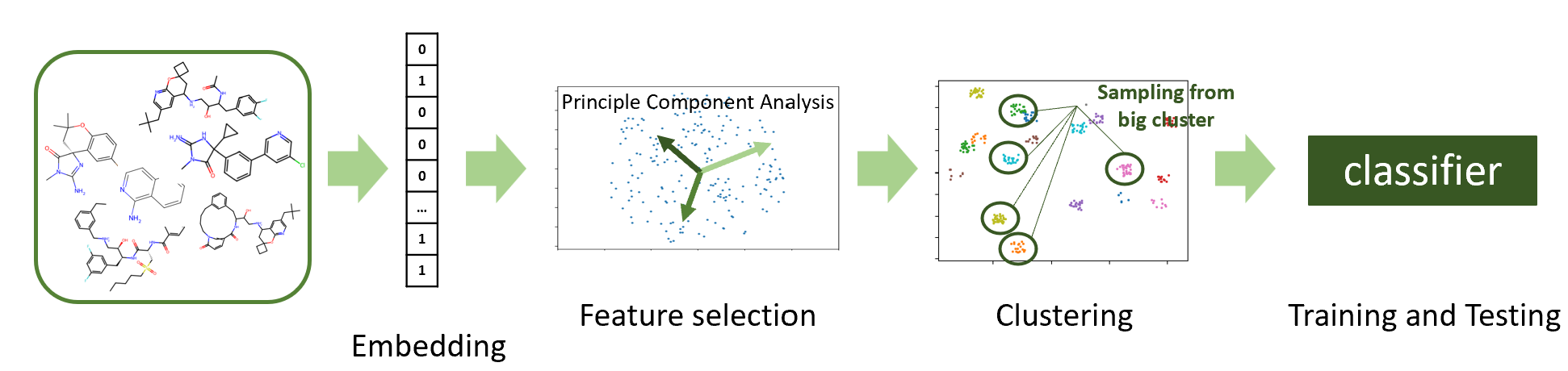}%{Figures/cluster_sampling.svg}
\caption{To test the generalization ability of classical and quantum models, cluster\cite{butina1999unsupervised} of embedded molecules was conducted, and 1 to 7 data points from each larger cluster were selected as training data. (embedding method: MGFP)}
\label{fig:clustering_sampling}
\end{figure}

\begin{figure}[h!]
    \centering
    \includegraphics[width=0.6\linewidth]{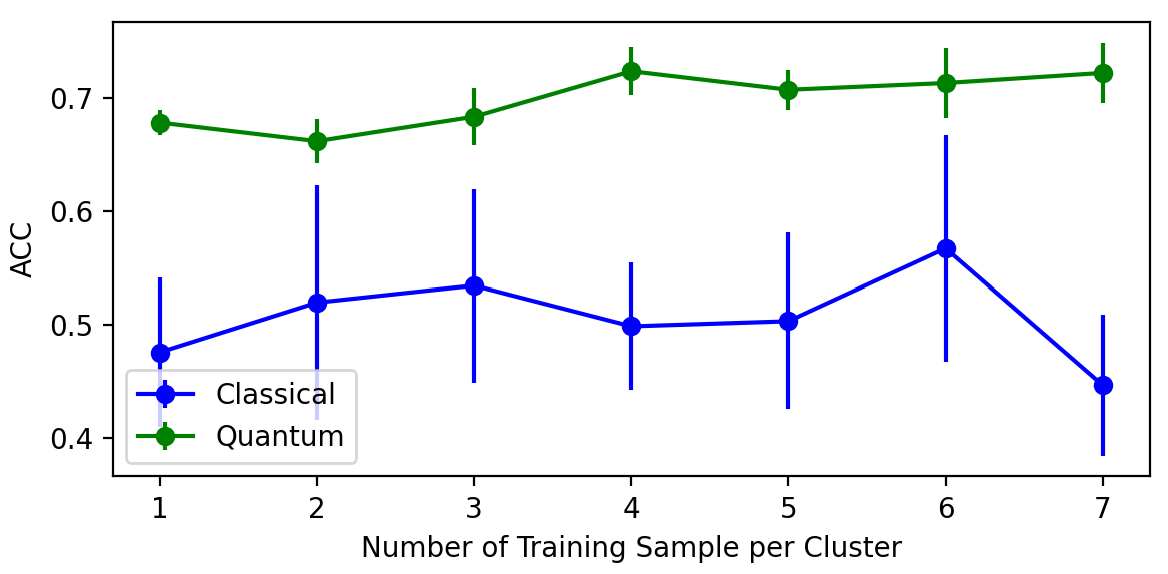}
    \caption{The comparisons of the testing accuracies when different numbers of samples picked from each cluster are used for training. In the scenario of low-dimensional features, $n=3$, the testing accuracies are significantly higher in the quantum classifier.}
    \label{fig:MGF_BACE_cluster}
\end{figure}

The generalization abilities of classical/quantum classifiers in low-dimensional feature regimes are also identified. One can accurately control the training data properties and ensure that similar features have been seen by the model by selecting the training sample from clustered data. The training data is clustered, followed by the training sample pick-up from the cluster with a size larger than $20$. The number of training samples from different clusters is varied from $1-7$ (Fig.\ref{fig:clustering_sampling}). From the Fig.\ref{fig:MGF_BACE_cluster}, we find the testing accuracies of the quantum classifier are higher even with the small number of training data, implying the better generalization ability of the quantum classifier for low-dimensional feature regime.

\subsection*{Generality of the quantum advantage on other datasets}
\begin{figure}[h!]
    \centering
    \includegraphics[width=0.6\linewidth]{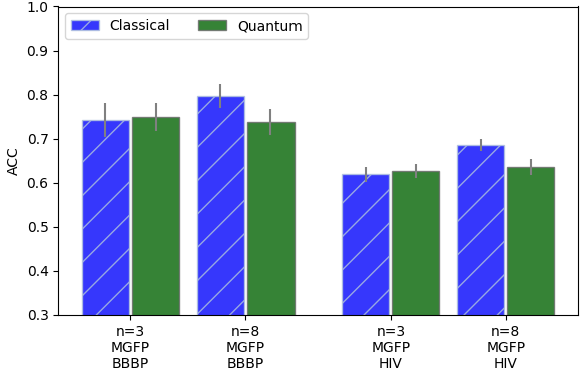}
    \caption{Performances of the classical and quantum classifiers: The histogram depicts the prediction accuracies under MGFP embedding for both $n=3$ and $n=8$. From the figure, it is evident that the quantum classifier performs slightly better than the classical one when $n$ is small. (The BBBP and HIV datasets are embedded using MGFP.))}
    \label{fig:BBBP_HIV_MGFP}
\end{figure}

\begin{figure}[h!]
    \centering
    \includegraphics[width=0.6\linewidth]{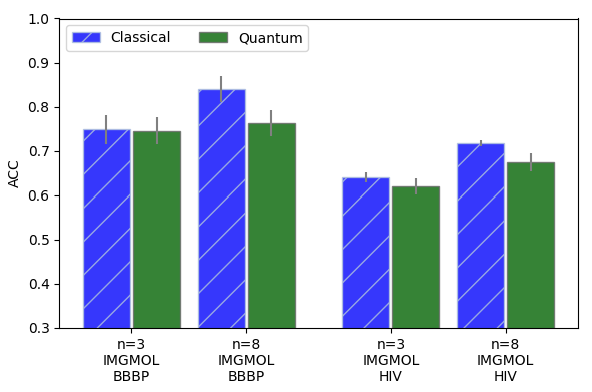}
    \caption{Performances of the classical/quantum classifiers. The histogram of the prediction accuracies under IMAGMOL embedding when $n=3$ and $n=8$. Data embedded by IMGMOL enhances the performances of classical classifiers but quantum classifiers. This improvement vanishes the slightly visible quantum advantages shown in Fig.\ref{fig:BBBP_HIV_MGFP}. }
    \label{fig:BBBP_HIV_IMGMOL}
\end{figure}

We tested the generality of this quantum advantage on different open datasets, HIV and BBBP. To address data imbalance issues, we undersampled the data before the classification procedure. As shown in Fig. \ref{fig:BBBP_HIV_MGFP}, classical classifiers exhibit better performance on both datasets when $n=8$ embedded by MGFP. However, when $n=3$, quantum classifiers perform slightly better by approximately 1\% and 2\% on BBBP and HIV datasets, respectively. The performance enhancement is not particularly pronounced. However, when the data is embedded using IMGMOL, the quantum advantages become less apparent due to the substantial improvement in the performance of classical classifiers (Fig. \ref{fig:BBBP_HIV_IMGMOL}). From Fig. \ref{fig:BBBP_HIV_MGFP} and Fig. \ref{fig:BBBP_HIV_IMGMOL}, it is evident that the performances of classical classifiers depend on the embedding methods and the number of features.

\section*{Discussion}

The application of machine learning in healthcare has been widely discussed; however, the model's performance strongly relies on data availability and quality. Additionally, the availability and completeness of features are common challenges during the collection of medical-related data, necessitating improved classification performance under these difficult circumstances. On the other hand, PQC-based models are known for their superior expressive power compared to classical MLP models. This quantum advantage can play a critical role in meeting the aforementioned demands. To explore this possibility, we initially apply classical and quantum classifiers with the most general architecture to classify the BACE dataset, which is embedded by MGFP/IMGMOL and followed by PCA feature selection. We find that the quantum classifier outperforms the classical one when dealing with incomplete feature data, and this advantage holds for different embedding methods in our experiments.
Furthermore, we delve into the effect of the number of training samples on each classifier's performance and observe that the quantum classifier performs better, especially when the number of available features is small. This observation allows us to leverage a PQC-based classifier on incomplete data, addressing data-related challenges.

To gain a deeper understanding of this quantum advantage, we systematically investigate the impact of feature numbers on the classifiers. We find that the classical classifier is more sensitive to the changes in feature numbers, while the quantum classifier exhibits robustness. This is evident in the relation plot of testing accuracy and training loss under different feature numbers ($n=3$ and $n=4$). While changes in feature numbers do not significantly impact the distribution in the quantum classifier, there are substantial changes in the distribution for classical classifiers with varying feature numbers(see Fig.\ref{fig:MGFP_ACC_n4_n8}(b)(c)). The robustness of the quantum classifier is also evident in its sensitivity to the number of training samples. This robustness results in slower degradation of testing accuracy concerning the number of training samples and highlights the quantum advantage in scenarios with a few features. Additionally, the relation plot demonstrates higher testing accuracy in the quantum classifier when the training loss is the same at $n=3$, implying better generalization power in quantum classifiers. When $n=4$, similar testing accuracies are achieved for both classical and quantum classifiers, while the quantum classifier has 18\% fewer trainable parameters, indicating superior expressive power.

The improved performance of the quantum classifier can be attributed to the limitation imposed by the VC dimension on generalization error. The generalization error bound is known to be influenced by the square root of the VC dimension divided by the sample number. For models composed of multi-layer perceptrons in classical classifiers, the VC dimension scales as $\varpropto \rho^2$\cite{bartlett2003vapnik}, where $\rho$ represents the number of trainable parameters. In contrast, according to quantum learning theory, the VC dimension of the circuit scales as $\varpropto \Gamma \log_2 \Gamma$\cite{dissertation}, where $\Gamma$ is the number of gates with trainable rotational angles. Our findings suggest that the model complexity increases more rapidly with the number of trainable parameters in classical models compared to quantum circuit models. This implies a tighter generalization error bound in quantum circuit models, resulting in improved generalization power.

Although Ref. \cite{1mensa2023quantum} and Ref. \cite{bhatia2023quantum} demonstrate that the quantum support vector classifier (QSVC) outperforms the classical one in terms of AUC-ROC score when data is represented by descriptors selected by PCA, their results further highlight that quantum advantages are more pronounced on some datasets with a higher number of features. In their work, the quantum algorithm is not PQC-based, and the number of features is limited to eight due to the encoding method in the circuit and the availability of qubits on NISQ devices. The QSVC comprises a quantum kernel and a classical classifier, showcasing the potential of quantum SVC as the available qubit count increases. 

In our experiments, the generalizability of the quantum advantage to other datasets(BBBP/HIV) is not significant from the statistical results; however, individual trials in the experiment still reveal the quantum advantage in specific data partitions with a small number of features, particularly in MGFP embedding (see \ref{fig:HIV_BBBP_MGFP} in Appendix A). This data dependency aligns with what is found in Ref. \cite{bhatia2023quantum}, where the quantum advantage does not manifest across all testing datasets. Investigating this dependency is intriguing but falls beyond the scope of this manuscript.

In this work, we aim to provide a fair comparison of the performances of classical and quantum classifiers on datasets with limited samples and features, showcasing the potential of quantum algorithms to address common issues in medical-related datasets. Although we have considered the most concise architecture for these classifiers in this study, the circuit/neural layers we have explored are general and can be applied to various other model structures. By understanding the inherent properties of these classifiers under different scenarios, we can effectively utilize circuit modules to implement larger-scale quantum machine learning models.

\section*{Conclusion}

In this study, we conducted a systematic investigation of the performance of classical and quantum classifiers on various datasets. We observed that the quantum classifier outperforms the classical one, particularly when the number of features and training samples is limited. The robustness of the quantum classifier further underscores this quantum advantage, whereas the classical classifier is more sensitive to changes in feature and sample numbers. Additionally, we examined the relationship between training loss and testing accuracy and found that the quantum classifier exhibits better expressive and generalization power when working with incomplete data. This shines the light on dealing with the learning problem of missing or incomplete data.

\section*{Dataset and Evaluation Metrics}

\subsection*{Dataset}
%BACE/BBBP/HIV/Clintox
\subsubsection*{BACE}
The BACE dataset\cite{subramanian2016computational} is composed of the binding results of 1522 inhibitory compounds to human beta-secretase 1 (BACE-1), which is important for myelin sheaths of neurons and is related to Alzheimer's disease. The inhibitors block the BACE, preventing the buildup of beta-amyloid and slowing down Alzheimer's disease. The raw data consists of SMILES of inhibitors, $pIC_{50}$ and class (bound or non-bound).

\subsubsection*{BBBP}
The blood-brain-barrier penetration (BBBP) dataset\cite{martins2012bayesian} collects 2053 molecules from the BBBP-related articles. The molecules are divided into two categories (penetrable/non-penetrable) according to the penetration ability. The raw data comprises molecule names, a simplified molecular-input line-entry system (SMILES), and binary labels for penetration/non-penetration.

\subsubsection*{HIV}
Drug Therapeutics Program (DTP) AIDS Antiviral Screen\cite{HIVdataset} introduced the HIV dataset consisting of \~ 40000 compounds SMILES labeled by confirmed inactive, confirmed active, and confirmed moderately active according to the ability to inhibit HIV replication. Three catalogs are further aggregated into inactive/active by combining confirmed active and confirmed moderately active into the active catalog such that the dataset can be used for QSAR prediction.

%\subsubsection*{Clintox}
%The Clintox data set consists of 1491 drugs which are either from the FDA-approved list\cite{novick2013sweetlead} or the list of drugs that failed the clinic trials for toxicity reasons\cite{aact2020aggregate}. The raw data is composed of the SMILES, FDA approval status and the clinic trial results.

\subsection*{Evaluation Metrics}
%accuracy/recall
The performance of the classifiers is evaluated by the accuracy score and recall score. The accuracy score is defined by:\\
$ACC = \frac{1}{n_{samples}}\sum^{n_{samples}-1}_{i=1}1(\hat{y}_i=y_{i})$\\
where $1(x)$ is the indicator function, $\hat{y}_i $ is the predicted value and $y_i$ is the true value. Recall score, also known as sensitivity, is defined by the ratio of $TP$ and positive samples and represented by $\frac{TP}{TP+FN}$. Where $TP$ and $FN$ are the number of true positives and false negatives, respectively.
To evaluate the performances moderately, the batched training data is shared by classical and quantum classifiers simultaneously in the training phase. The accuracy/recall scores are used to evaluate the same testing set. 

\bibliography{sample}

\begin{thebibliography}{10}
\urlstyle{rm}
\expandafter\ifx\csname url\endcsname\relax
  \def\url#1{\texttt{#1}}\fi
\expandafter\ifx\csname urlprefix\endcsname\relax\def\urlprefix{URL }\fi
\expandafter\ifx\csname doiprefix\endcsname\relax\def\doiprefix{DOI: }\fi
\providecommand{\bibinfo}[2]{#2}
\providecommand{\eprint}[2][]{\url{#2}}

\bibitem{hansch1964p}
\bibinfo{author}{Hansch, C.} \& \bibinfo{author}{Fujita, T.}
\newblock \bibinfo{journal}{\bibinfo{title}{p-$\sigma$-$\pi$ analysis. a method
  for the correlation of biological activity and chemical structure}}.
\newblock {\emph{\JournalTitle{Journal of the American Chemical Society}}}
  \textbf{\bibinfo{volume}{86}}, \bibinfo{pages}{1616--1626}
  (\bibinfo{year}{1964}).

\bibitem{zhou2006boosting}
\bibinfo{author}{Zhou, Y.-P.} \emph{et~al.}
\newblock \bibinfo{journal}{\bibinfo{title}{Boosting support vector regression
  in qsar studies of bioactivities of chemical compounds}}.
\newblock {\emph{\JournalTitle{European journal of pharmaceutical sciences}}}
  \textbf{\bibinfo{volume}{28}}, \bibinfo{pages}{344--353}
  (\bibinfo{year}{2006}).

\bibitem{pourbasheer2009application}
\bibinfo{author}{Pourbasheer, E.}, \bibinfo{author}{Riahi, S.},
  \bibinfo{author}{Ganjali, M.~R.} \& \bibinfo{author}{Norouzi, P.}
\newblock \bibinfo{journal}{\bibinfo{title}{Application of genetic
  algorithm-support vector machine (ga-svm) for prediction of bk-channels
  activity}}.
\newblock {\emph{\JournalTitle{European journal of medicinal chemistry}}}
  \textbf{\bibinfo{volume}{44}}, \bibinfo{pages}{5023--5028}
  (\bibinfo{year}{2009}).

\bibitem{polishchuk2009application}
\bibinfo{author}{Polishchuk, P.~G.} \emph{et~al.}
\newblock \bibinfo{journal}{\bibinfo{title}{Application of random forest
  approach to qsar prediction of aquatic toxicity}}.
\newblock {\emph{\JournalTitle{Journal of chemical information and modeling}}}
  \textbf{\bibinfo{volume}{49}}, \bibinfo{pages}{2481--2488}
  (\bibinfo{year}{2009}).

\bibitem{dahl2014multi}
\bibinfo{author}{Dahl, G.~E.}, \bibinfo{author}{Jaitly, N.} \&
  \bibinfo{author}{Salakhutdinov, R.}
\newblock \bibinfo{journal}{\bibinfo{title}{Multi-task neural networks for qsar
  predictions}}.
\newblock {\emph{\JournalTitle{arXiv preprint arXiv:1406.1231}}}
  (\bibinfo{year}{2014}).

\bibitem{kwon2019comprehensive}
\bibinfo{author}{Kwon, S.}, \bibinfo{author}{Bae, H.}, \bibinfo{author}{Jo, J.}
  \& \bibinfo{author}{Yoon, S.}
\newblock \bibinfo{journal}{\bibinfo{title}{Comprehensive ensemble in qsar
  prediction for drug discovery}}.
\newblock {\emph{\JournalTitle{BMC bioinformatics}}}
  \textbf{\bibinfo{volume}{20}}, \bibinfo{pages}{1--12} (\bibinfo{year}{2019}).

\bibitem{pourbasheer2019qsar}
\bibinfo{author}{Pourbasheer, E.}, \bibinfo{author}{Aalizadeh, R.} \&
  \bibinfo{author}{Ganjali, M.~R.}
\newblock \bibinfo{journal}{\bibinfo{title}{Qsar study of ck2 inhibitors by
  ga-mlr and ga-svm methods}}.
\newblock {\emph{\JournalTitle{Arabian Journal of Chemistry}}}
  \textbf{\bibinfo{volume}{12}}, \bibinfo{pages}{2141--2149}
  (\bibinfo{year}{2019}).

\bibitem{hu2020deep}
\bibinfo{author}{Hu, S.}, \bibinfo{author}{Chen, P.}, \bibinfo{author}{Gu, P.}
  \& \bibinfo{author}{Wang, B.}
\newblock \bibinfo{journal}{\bibinfo{title}{A deep learning-based chemical
  system for qsar prediction}}.
\newblock {\emph{\JournalTitle{IEEE journal of biomedical and health
  informatics}}} \textbf{\bibinfo{volume}{24}}, \bibinfo{pages}{3020--3028}
  (\bibinfo{year}{2020}).

\bibitem{kim2020extension}
\bibinfo{author}{Kim, B.~C.}, \bibinfo{author}{Joe, D.}, \bibinfo{author}{Woo,
  Y.}, \bibinfo{author}{Kim, Y.} \& \bibinfo{author}{Yoon, G.}
\newblock \bibinfo{journal}{\bibinfo{title}{Extension of pqsar: Ensemble model
  generated by random forest and partial least squares regressions}}.
\newblock {\emph{\JournalTitle{IEEE Access}}} \textbf{\bibinfo{volume}{8}},
  \bibinfo{pages}{180087--180099} (\bibinfo{year}{2020}).

\bibitem{11benedetti2019parameterized}
\bibinfo{author}{Benedetti, M.}, \bibinfo{author}{Lloyd, E.},
  \bibinfo{author}{Sack, S.} \& \bibinfo{author}{Fiorentini, M.}
\newblock \bibinfo{journal}{\bibinfo{title}{Parameterized quantum circuits as
  machine learning models}}.
\newblock {\emph{\JournalTitle{Quantum Science and Technology}}}
  \textbf{\bibinfo{volume}{4}}, \bibinfo{pages}{043001} (\bibinfo{year}{2019}).

\bibitem{15biamonte2017quantum}
\bibinfo{author}{Biamonte, J.} \emph{et~al.}
\newblock \bibinfo{journal}{\bibinfo{title}{Quantum machine learning}}.
\newblock {\emph{\JournalTitle{Nature}}} \textbf{\bibinfo{volume}{549}},
  \bibinfo{pages}{195--202} (\bibinfo{year}{2017}).

\bibitem{16tsang2022hybrid}
\bibinfo{author}{Tsang, S.~L.}, \bibinfo{author}{West, M.~T.},
  \bibinfo{author}{Erfani, S.~M.} \& \bibinfo{author}{Usman, M.}
\newblock \bibinfo{journal}{\bibinfo{title}{Hybrid quantum-classical generative
  adversarial network for high resolution image generation}}.
\newblock {\emph{\JournalTitle{arXiv preprint arXiv:2212.11614}}}
  (\bibinfo{year}{2022}).

\bibitem{17huang2021experimental}
\bibinfo{author}{Huang, H.-L.} \emph{et~al.}
\newblock \bibinfo{journal}{\bibinfo{title}{Experimental quantum generative
  adversarial networks for image generation}}.
\newblock {\emph{\JournalTitle{Physical Review Applied}}}
  \textbf{\bibinfo{volume}{16}}, \bibinfo{pages}{024051}
  (\bibinfo{year}{2021}).

\bibitem{4suzuki2020predicting}
\bibinfo{author}{Suzuki, T.} \& \bibinfo{author}{Katouda, M.}
\newblock \bibinfo{journal}{\bibinfo{title}{Predicting toxicity by quantum
  machine learning}}.
\newblock {\emph{\JournalTitle{Journal of Physics Communications}}}
  \textbf{\bibinfo{volume}{4}}, \bibinfo{pages}{125012} (\bibinfo{year}{2020}).

\bibitem{10wu2021application}
\bibinfo{author}{Wu, S.~L.} \emph{et~al.}
\newblock \bibinfo{journal}{\bibinfo{title}{Application of quantum machine
  learning using the quantum kernel algorithm on high energy physics analysis
  at the lhc}}.
\newblock {\emph{\JournalTitle{Physical Review Research}}}
  \textbf{\bibinfo{volume}{3}}, \bibinfo{pages}{033221} (\bibinfo{year}{2021}).

\bibitem{14havlivcek2019supervised}
\bibinfo{author}{Havl{\'\i}{\v{c}}ek, V.} \emph{et~al.}
\newblock \bibinfo{journal}{\bibinfo{title}{Supervised learning with
  quantum-enhanced feature spaces}}.
\newblock {\emph{\JournalTitle{Nature}}} \textbf{\bibinfo{volume}{567}},
  \bibinfo{pages}{209--212} (\bibinfo{year}{2019}).

\bibitem{7batra2021quantum}
\bibinfo{author}{Batra, K.} \emph{et~al.}
\newblock \bibinfo{journal}{\bibinfo{title}{Quantum machine learning algorithms
  for drug discovery applications}}.
\newblock {\emph{\JournalTitle{Journal of chemical information and modeling}}}
  \textbf{\bibinfo{volume}{61}}, \bibinfo{pages}{2641--2647}
  (\bibinfo{year}{2021}).

\bibitem{8liu2021rigorous}
\bibinfo{author}{Liu, Y.}, \bibinfo{author}{Arunachalam, S.} \&
  \bibinfo{author}{Temme, K.}
\newblock \bibinfo{journal}{\bibinfo{title}{A rigorous and robust quantum
  speed-up in supervised machine learning}}.
\newblock {\emph{\JournalTitle{Nature Physics}}} \textbf{\bibinfo{volume}{17}},
  \bibinfo{pages}{1013--1017} (\bibinfo{year}{2021}).

\bibitem{18du2020expressive}
\bibinfo{author}{Du, Y.}, \bibinfo{author}{Hsieh, M.-H.}, \bibinfo{author}{Liu,
  T.} \& \bibinfo{author}{Tao, D.}
\newblock \bibinfo{journal}{\bibinfo{title}{Expressive power of parametrized
  quantum circuits}}.
\newblock {\emph{\JournalTitle{Physical Review Research}}}
  \textbf{\bibinfo{volume}{2}}, \bibinfo{pages}{033125} (\bibinfo{year}{2020}).

\bibitem{12cao2018potential}
\bibinfo{author}{Cao, Y.}, \bibinfo{author}{Romero, J.} \&
  \bibinfo{author}{Aspuru-Guzik, A.}
\newblock \bibinfo{journal}{\bibinfo{title}{Potential of quantum computing for
  drug discovery}}.
\newblock {\emph{\JournalTitle{IBM Journal of Research and Development}}}
  \textbf{\bibinfo{volume}{62}}, \bibinfo{pages}{6--1} (\bibinfo{year}{2018}).

\bibitem{5kao2023exploring}
\bibinfo{author}{Kao, P.-Y.} \emph{et~al.}
\newblock \bibinfo{journal}{\bibinfo{title}{Exploring the advantages of quantum
  generative adversarial networks in generative chemistry}}.
\newblock {\emph{\JournalTitle{Journal of Chemical Information and Modeling}}}
  (\bibinfo{year}{2023}).

\bibitem{3domingo2023hybrid}
\bibinfo{author}{Domingo, L.}, \bibinfo{author}{Djukic, M.},
  \bibinfo{author}{Johnson, C.} \& \bibinfo{author}{Borondo, F.}
\newblock \bibinfo{journal}{\bibinfo{title}{Hybrid quantum-classical
  convolutional neural networks to improve molecular protein binding affinity
  predictions}}.
\newblock {\emph{\JournalTitle{arXiv preprint arXiv:2301.06331}}}
  (\bibinfo{year}{2023}).

\bibitem{2sagingalieva2023hybrid}
\bibinfo{author}{Sagingalieva, A.} \emph{et~al.}
\newblock \bibinfo{journal}{\bibinfo{title}{Hybrid quantum neural network for
  drug response prediction}}.
\newblock {\emph{\JournalTitle{Cancers}}} \textbf{\bibinfo{volume}{15}},
  \bibinfo{pages}{2705} (\bibinfo{year}{2023}).

\bibitem{19hekler2019we}
\bibinfo{author}{Hekler, E.~B.} \emph{et~al.}
\newblock \bibinfo{journal}{\bibinfo{title}{Why we need a small data
  paradigm}}.
\newblock {\emph{\JournalTitle{BMC medicine}}} \textbf{\bibinfo{volume}{17}},
  \bibinfo{pages}{1--9} (\bibinfo{year}{2019}).

\bibitem{jolliffe2002principal}
\bibinfo{author}{Jolliffe, I.~T.}
\newblock \emph{\bibinfo{title}{Principal component analysis for special types
  of data}} (\bibinfo{publisher}{Springer}, \bibinfo{year}{2002}).

\bibitem{rogers2010extended}
\bibinfo{author}{Rogers, D.} \& \bibinfo{author}{Hahn, M.}
\newblock \bibinfo{journal}{\bibinfo{title}{Extended-connectivity
  fingerprints}}.
\newblock {\emph{\JournalTitle{Journal of chemical information and modeling}}}
  \textbf{\bibinfo{volume}{50}}, \bibinfo{pages}{742--754}
  (\bibinfo{year}{2010}).

\bibitem{zeng2022accurate}
\bibinfo{author}{Zeng, X.} \emph{et~al.}
\newblock \bibinfo{journal}{\bibinfo{title}{Accurate prediction of molecular
  properties and drug targets using a self-supervised image representation
  learning framework}}.
\newblock {\emph{\JournalTitle{Nature Machine Intelligence}}}
  \textbf{\bibinfo{volume}{4}}, \bibinfo{pages}{1004--1016}
  (\bibinfo{year}{2022}).

\bibitem{greg_landrum_2022_7179566}
\bibinfo{author}{Landrum, G.} \emph{et~al.}
\newblock \bibinfo{title}{rdkit/rdkit: 2022\_09\_1b1 (q3 2022) release},
  \doiprefix\url{10.5281/zenodo.7179566} (\bibinfo{year}{2022}).

\bibitem{butina1999unsupervised}
\bibinfo{author}{Butina, D.}
\newblock \bibinfo{journal}{\bibinfo{title}{Unsupervised data base clustering
  based on daylight's fingerprint and tanimoto similarity: A fast and automated
  way to cluster small and large data sets}}.
\newblock {\emph{\JournalTitle{Journal of Chemical Information and Computer
  Sciences}}} \textbf{\bibinfo{volume}{39}}, \bibinfo{pages}{747--750}
  (\bibinfo{year}{1999}).

\bibitem{bartlett2003vapnik}
\bibinfo{author}{Bartlett, P.~L.} \& \bibinfo{author}{Maass, W.}
\newblock \bibinfo{journal}{\bibinfo{title}{Vapnik-chervonenkis dimension of
  neural nets}}.
\newblock {\emph{\JournalTitle{The handbook of brain theory and neural
  networks}}} \bibinfo{pages}{1188--1192} (\bibinfo{year}{2003}).

\bibitem{dissertation}
\bibinfo{author}{Caro, M.~C.}
\newblock \emph{\bibinfo{title}{Quantum Learning Theory}}.
\newblock Ph.D. thesis, \bibinfo{school}{Technische Universität München}
  (\bibinfo{year}{2022}).

\bibitem{1mensa2023quantum}
\bibinfo{author}{Mensa, S.}, \bibinfo{author}{Sahin, E.},
  \bibinfo{author}{Tacchino, F.}, \bibinfo{author}{Barkoutsos, P.~K.} \&
  \bibinfo{author}{Tavernelli, I.}
\newblock \bibinfo{journal}{\bibinfo{title}{Quantum machine learning framework
  for virtual screening in drug discovery: a prospective quantum advantage}}.
\newblock {\emph{\JournalTitle{Machine Learning: Science and Technology}}}
  \textbf{\bibinfo{volume}{4}}, \bibinfo{pages}{015023} (\bibinfo{year}{2023}).

\bibitem{bhatia2023quantum}
\bibinfo{author}{Bhatia, A.~S.}, \bibinfo{author}{Saggi, M.~K.} \&
  \bibinfo{author}{Kais, S.}
\newblock \bibinfo{journal}{\bibinfo{title}{Quantum machine learning predicting
  adme-tox properties in drug discovery}}.
\newblock {\emph{\JournalTitle{Journal of Chemical Information and Modeling}}}
  (\bibinfo{year}{2023}).

\bibitem{subramanian2016computational}
\bibinfo{author}{Subramanian, G.}, \bibinfo{author}{Ramsundar, B.},
  \bibinfo{author}{Pande, V.} \& \bibinfo{author}{Denny, R.~A.}
\newblock \bibinfo{journal}{\bibinfo{title}{Computational modeling of
  $\beta$-secretase 1 (bace-1) inhibitors using ligand based approaches}}.
\newblock {\emph{\JournalTitle{Journal of chemical information and modeling}}}
  \textbf{\bibinfo{volume}{56}}, \bibinfo{pages}{1936--1949}
  (\bibinfo{year}{2016}).

\bibitem{martins2012bayesian}
\bibinfo{author}{Martins, I.~F.}, \bibinfo{author}{Teixeira, A.~L.},
  \bibinfo{author}{Pinheiro, L.} \& \bibinfo{author}{Falcao, A.~O.}
\newblock \bibinfo{journal}{\bibinfo{title}{A bayesian approach to in silico
  blood-brain barrier penetration modeling}}.
\newblock {\emph{\JournalTitle{Journal of chemical information and modeling}}}
  \textbf{\bibinfo{volume}{52}}, \bibinfo{pages}{1686--1697}
  (\bibinfo{year}{2012}).

\bibitem{HIVdataset}
\bibinfo{title}{Aids antiviral screen data}.
\newblock
  \bibinfo{howpublished}{\url{https://wiki.nci.nih.gov/display/NCIDTPdata/AIDS+Antiviral+Screen+Data}}.
\newblock \bibinfo{note}{Accessed: 27.09.2017}.

\end{thebibliography}

%\section*{Acknowledgements (not compulsory)}

%Wei-Yin Chiang thanks balabala for balabala

%\section*{Author contributions statement}

%Wei-Yin Chiang and Ya-Chu Yang designed and performed the experiments and wrote the article. Po-Yu Kao extracted the molecular embedding, wrote, and proofread the article. Yen-Chu Lin coordinated the project and reviewed the articles.

\section*{Appendix A}

\begin{figure}[h!]
    \centering
    \includegraphics[width=\linewidth]{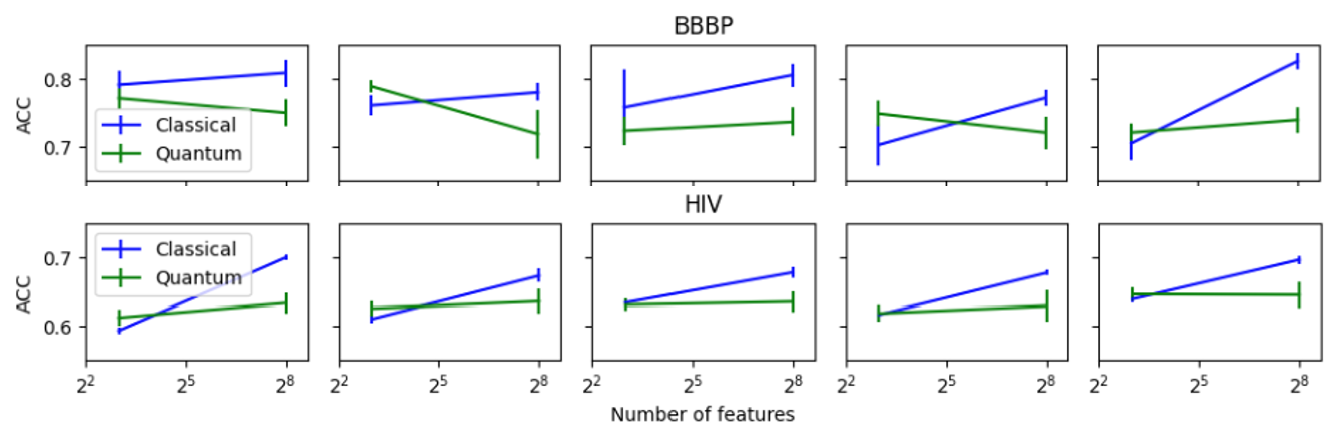}
    \caption{QSAR prediction on MGFP embedded BBBP(top) and HIV(bottom) datasets from classical(blue) and quantum(green) classifiers. The results in different columns are obtained from different data partitions. Quantum classifiers outperform the classical ones in low-dimensional feature regimes in most data partitions. This partition-dependent behavior results in an insignificant quantum advantage.}
    \label{fig:HIV_BBBP_MGFP}
\end{figure}

The results shown in the main text demonstrate the insignificant quantum advantage on the BBBP and HIV datasets. This insignificance results from the various behavior of classifiers on different data partitions and is shown in Fig.\ref{fig:HIV_BBBP_MGFP}.

A similar task can also apply to the data embedded by different methods. The quantum advantage disappears when the BBBP/HIV data are embedded by IMGMOL. Similar to the partition-dependent results in the previous demonstrations, however, the classification accuracy of the quantum classifier is wore in most cases.
\begin{figure}[h!]
    \centering
    \includegraphics[width=\linewidth]{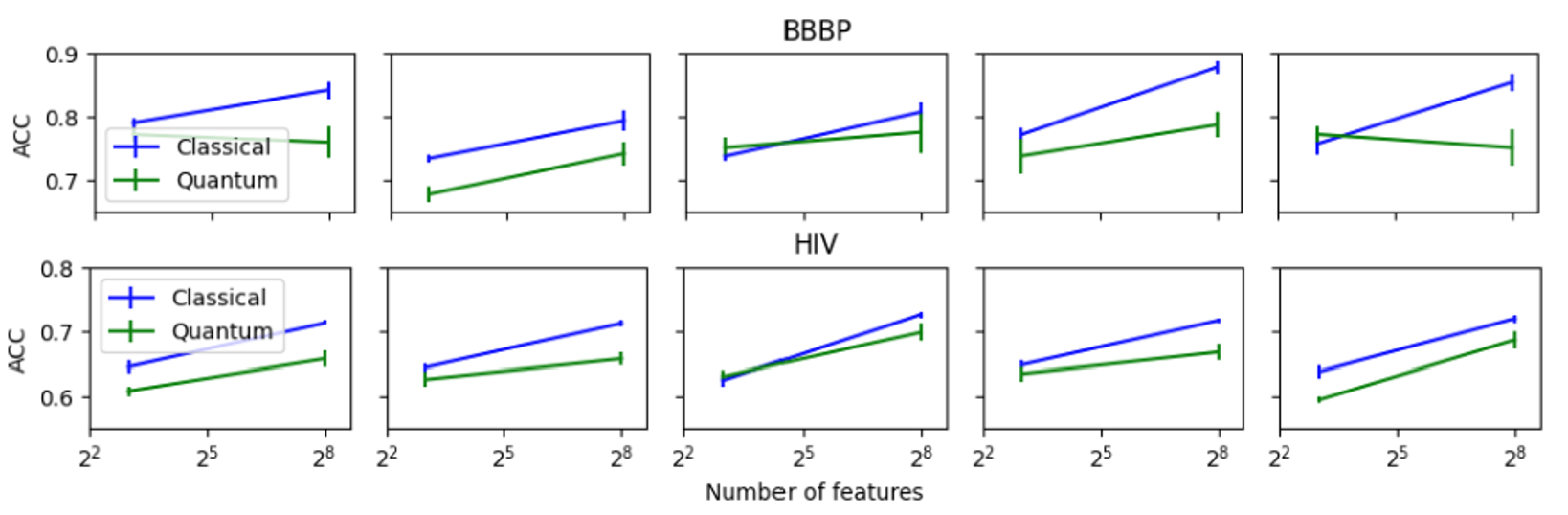}
    \caption{QSAR prediction on IMGMOL embedded BBBP(top) and HIV(bottom) datasets from classical(blue) and quantum(green) classifiers. The results in different columns are obtained from different data partitions. Quantum classifiers outperform the classical ones in low-dimensional feature regimes only in a few data partitions. Most partitions do not show quantum advantage in classification.}
    \label{fig:HIV_BBBP_IMAGMOL}
\end{figure}
\end{document}